\documentclass[12pt]{article}

\usepackage[utf8]{inputenc}
\usepackage[T1]{fontenc}
\usepackage{hyperref}
\usepackage{url}
\usepackage{booktabs}
\usepackage{amssymb}
\usepackage{amsfonts}
\usepackage{nicefrac}
\usepackage{microtype}
\usepackage{lipsum}
\usepackage{graphicx}
\usepackage{tcolorbox}
\tcbuselibrary{breakable, skins, listings}
\usepackage{float}
\usepackage{amsmath}
\usepackage{enumitem}
\graphicspath{{./images/}}
\usepackage{xcolor}
\usepackage{listings}

\title{BioModelsRAG: A Biological Modeling Assistant Using RAG (Retrieval Augmented Generation)}

\author{%
Bhavyahshree Navaneetha Krishnan\thanks{University of Washington, Seattle, WA 98195. Email: \href{mailto:bhavyak7@uw.edu}{bhavyak7@uw.edu}} \and
Adel Heydarabadipour\thanks{Department of Bioengineering, University of Washington, Seattle, WA 98195. Email: \href{mailto:adelhp@uw.edu}{adelhp@uw.edu}} \and
Herbert Sauro\thanks{Department of Bioengineering, University of Washington, Seattle, WA 98195. Email: \href{mailto:hsauro@uw.edu}{hsauro@uw.edu}}
}

\date{} 

\begin{document}

\maketitle

\begin{abstract}
The BioModels database is one of the premier databases for computational models in systems biology. The database contains over 1000 curated models and an even larger number of non-curated models. All the models are stored in the machine-readable format, SBML. Although SBML can be translated into the human readable Antimony format, analyzing the models can still be time consuming. 

In order to bridge this gap, a LLM (large language model) assistant was created to analyze the BioModels and allow interaction between the user and the model using natural language. By doing so, a user can easily and rapidly extract the salient points in a given model. Our analysis workflow involved `chunking' BioModels and converting them to plain text using llama3, and then embedding them in a ChromaDB database. The user-provided query was also embedded, and a similarity search was performed between the query and the BioModels in ChromaDB to extract the most relevant BioModels. The BioModels were then used as context to create the most accurate output in the chat between the user and the LLM. This approach greatly minimized the chance of hallucination and kept the LLM focused on the problem at hand. 

We illustrate the utility of this approach with a number of examples. The code is available at \url{https://github.com/TheBobBob/BioModelsRAG}. The website implementation is available at \url{https://biomodelsrag.streamlit.app/}.
\end{abstract}

\textbf{Keywords:} BioModels, LLMs, Retrieval Augmented Generation, Systems Biology

\section{Introduction}
BioModels is a repository of mathematical models of biological and biomedical systems. It hosts a large selection of existing literature-based physiologically and pharmaceutically relevant mechanistic models in standard formats such as SBML \cite{biomodelswebsite}\cite{Keating2020}. However, even with their associated published paper, these models can be hard to understand without explicit understanding of the biological mechanism described in the model. This  limits the usage of these model for biologists without background information in systems biology modeling. In addition, the BioModels website does not provide the more human-readable version of SBML in the form of Antimony \cite{smith2009antimony}. 

LLMs or \textbf{L}arge \textbf{L}anguage \textbf{M}odels include models such as ChatGPT, Claude, and Gemini \cite{banerjee}. These models function by using deep learning architectures, trained on massive datasets from the internet, such as books, and the GitHub repository. In generating code and text, LLMs have shown great potential \cite{codeanalysisllms}. Therefore, LLMs have great possibilities in analyzing Antimony models, specifically in converting Antimony into even simpler human-readable text to facilitate understanding by the user.

Retrieval Augmented Generation (RAG) is a technique that can be used to reduce hallucinations in LLMs. Hallucinations often occur in LLMs, and this is due to the fact that when LLMs are unable to produce coherent output, they fill in the blanks by including data that is plausible, but incorrect, especially while performing knowledge-intensive Natural Language Processing (NLP) tasks \cite{banerjee}\cite{louispatrick2024}. RAG can be used to provide relevant context to the LLM, so it is able to draw information from the context and fill in the hyper-specific information not present in its training data, and prevent hallucination of the model, which could be detrimental to the user's understanding. In this study, RAG was used to implement a chatbot where the user can query for different BioModels and subsequently "converse" with the LLM about a specific BioModel to improve the user's understanding without the risk of hallucinations from general LLMs.

\section{Methods}

In alignment with the objectives highlighted in the introduction, a Retrieval-Augmented Generation (RAG) based system was developed using Large Language Models (LLMs) to provide a user-friendly interface for individuals to dynamically interact with biomodels, even those without extensive backgrounds in systems biology or computational modeling. The chatbot was also deployed as a web application, enabling users without a command line interface or an IDE set up to use the chatbot through the web. 

In order to construct the RAG assistant, a multi-part pipeline was created. The first part of the pipeline is preparing and caching the models on a vector database. The second part of the pipeline is retrieving the relevant context from this database using user's query, putting it into the prompt, passing into the LLM and returning the LLM's response to the user. The following is the detail of each step of the pipeline.

\subsection{Downloading, converting, and segmenting BioModels}
All biomodels were downloaded via an API connecting the chatbot's backend to a \href{https://github.com/TheBobBob/BiomodelsCache}{GitHub repository} with the SBML version of the models. Using the Tellurium and Antimony libraries, all models were converted from SBML format to Antimony format to enable easier segmentation \cite{Keating2020} \cite{smith2024updates}. 

\subsection{Conversion of the Antimony model into plain text}
Through some preliminary experiments, we noticed that when the entire Antimony model was provided to the LLM and a simple query was asked, the LLM failed to accurately answer the question. We deduced that this was due to the fact that providing the LLM with the entire model caused it to overload the LLM's context window. In order to prevent this, the model was split into its constituents. 

The models were split into the following sections using text-based splitting: 
\begin{itemize}
    \item Compartments and species 
    \item Assignment rules 
    \item Reactions 
    \item Species initializations 
    \item Compartment initializations 
    \item Variable initializations 
    \item Other declarations 
    \item Unit definitions 
    \item Display names 
    \item CV terms
\end{itemize}

In order to split the biomodel, certain text patterns were searched for in the model that indicated the beginning of a new section. 

Each segment of Antimony was passed into an LLM with the goal of summarizing the segment while maintaining the key information to convert it into plain text to improve the accuracy of the final response. 

\lstset{
  basicstyle=\ttfamily\small,
  backgroundcolor=\color{gray!10},
  frame=single,
  breaklines=true,
  numbers=none,
  captionpos=b
}

The prompt used for summarization is seen below in Listing~\ref{lst:prompt}. The prompt was chosen in such a way to be as direct as possible to the LLM to prevent any misunderstandings. We codified the general principles that produced the most optimal output and embedded them in the prompt.

\begin{lstlisting}[caption={Prompt used for summarization}, label={lst:prompt}]
Summarize the following segment of Antimony in a clear and concise manner:

1. Provide a detailed summary using a limited number of words.
2. Maintain all original values and include any mathematical expressions or values in full.
3. Ensure that all variable names and their values are clearly presented.
4. Write the summary in paragraph format, putting an emphasis on clarity and completeness.

Here is the Antimony segment to summarize: 
\end{lstlisting}

\subsection{Storage of the Antimony models in a vector database}
Embedding is the process of converting unstructured data (such as text) into a numerical representation that computers can understand. The text chunks are projected as vectors into a multi dimensional space, which allows them to be represented as numbers \cite{vector}. Each summarized segment was added to a ChromaDB (a database hosted by the organization Chroma), to be queried with the user's prompt later on so that the user can get the most relevant pieces of context in their response. Embedding was conducted using the sentence-transformers package, specifically the all-MiniLM-L6-v2 model. The all-Mini-LM-L6-v2 model is better suited for this task compared to earlier models as it is a very lightweight model that processes queries quicker than previous models. The ChromaDB was initialized through the web service so that once every biomodel was processed, they could be queried from the database without needing to regenerate the summaries.

\subsection{Receiving user's query}
The user's query is received from the frontend of Streamlit through a text entry box. Once the query is received by Streamlit, it used to query the database in the next step.

\subsection{Database querying}
In order to find the most relevant parts of the biomodel that correlate with the query, a vector search is performed on the Chroma database initialized in the previous step. In order to perform the vector search, cosine similarity is calculated between the user-provided query and every query in the database. Cosine similarity is a representation of how closely "related" the two vectors are. It does this by calculating the cosine of the angle formed by the two vectors, once they are projected into a higher dimensional space. A higher cosine similarity indicates a higher correlation between the two text chunks. A database search is done and the 5 pieces of context with the highest cosine similarity are used as context with the final answer generation.

\subsection{Creation of prompt for final response}
The prompt used by the LLM to obtain the final response is shown below. The retrieved data from the database is added into the prompt as seen below (it is included in the text area labeled "context"). The prompt is fine-tuned for each specific query by providing the context - this allows the LLM to not hallucinate, as it is provided background context to answer the user's question with. The prompt contains the following: 
\begin{lstlisting}
- Context Piece 1 
- Context Piece 2 
- Context Piece 3 
- Context Piece 4 
- Context Piece 5 

Question from the User 
Note to LLM about telling the user when it does not have information on a query
\end{lstlisting}

\subsection{Generation of final response}
The final response to the question the user provided is generated by providing the RAG-LLM system with the above prompt. It contains the relevant context from the database querying and the user's question, so an accurate output can be generated.
\newline 

\begin{tcolorbox}[colback=blue!5!white, colframe=blue!75!black, title=Prompt to LLM for final response that is given to user]
You are a conversational agent. Now, using the context provided below

\begin{itemize}
    \item ${context}$
\end{itemize}

Answer the following question:
\begin{itemize}
    \item ${question}$
\end{itemize}

If the information is insufficient to answer the question, please state that clearly.
\end{tcolorbox}

\begin{figure}[tbp]
    \centering
    \includegraphics[width=\textwidth]{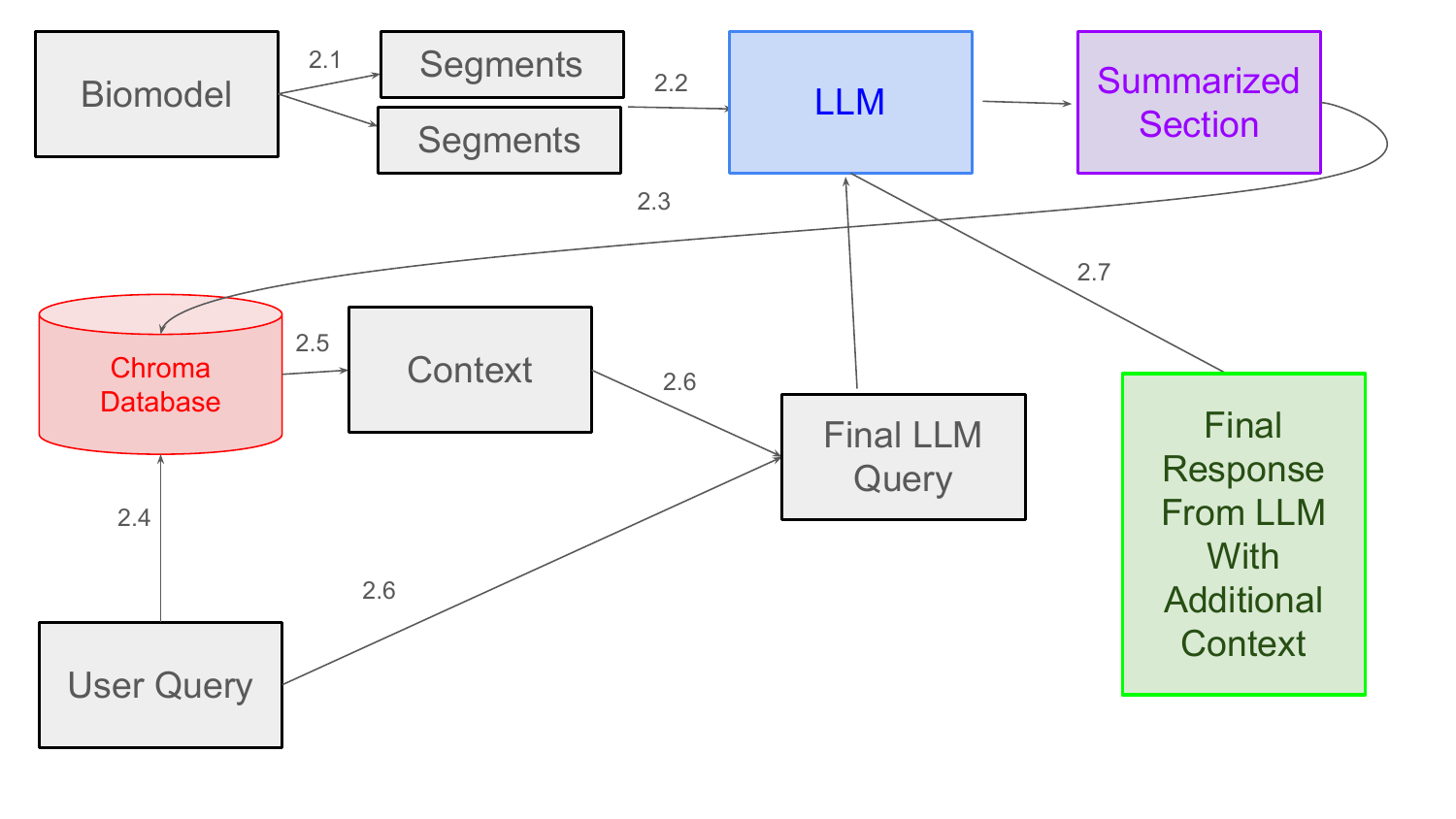}
    \caption{A visualization of the pipeline with the exact steps labeled corresponding to the section name}
    \label{fig:visualization of pipeline}
\end{figure}

\section{Results}
In order to test the RAG system's performance, two measures were used: fuzzy string matching (which is measured using token set ratio), and cosine similarity. These properties were used to test the following properties: whether the "main idea" is correct, whether the LLM "hallucinated" (or in layman's terms, "made up") the final answer, and whether the response is on-topic. 

Fuzzy string matching was used to determine how similar the "main idea" is between the text by removing common words like ``the'', ``and'', and others, and only retaining the words that add meaning to the text. When we compare the provided context, and the LLM output, we can determine whether the LLM is including words that are not referenced in the context, indicating that it could be hallucinating. If the token set ratio is higher (implemented through fuzzy string matching), there is increased similarity, indicating that it is more similar, and therefore containing less hallucination. Cosine similarity was used after it was established that the RAG sample contained less hallucination and relevant to the context (measured using the token set ratio). Cosine similarity measured how relevant the non-RAG sample was to the RAG sample, which determined whether the non-RAG sample was on-topic and actually answered the user's question. When token set ratio and cosine similarity were calculated, significant differences were found between the RAG and non-RAG samples.

To conduct these tests (as well as create the LLM that the local version of the chatbot was deployed with), the following custom parameters were found through preliminary tests. Custom parameters (temperature of 0, $top_{k}$ of 25 and $top_{p}$ of 1.5) improved the summarization output, by slightly increasing the token set ratio (slightly decreasing the likelihood for hallucination to be present in the summarization output). The $top_{p}$ is a parameter that controls the "randomness" of the model's output, the $top_{k}$ restricts the range of possible of words by reducing the part of the probability distribution that the LLM samples from when picking the next word, and the temperature is another parameter that controls the randomness of the output. The goal of these tests was to find the optimal set of parameters to prevent the LLM from making up information (hallucination). Through these tests, we discovered that a temperature of 0, and the provided $top_{p}$ of 1.0 and $top_{k}$ of 20 were the most optimal.

These results have shown that responses using the RAG technique are more detailed (as indicated by the higher token set ratio) and have reduced hallucination due to the fact that the LLM has the context necessary to output a scientifically accurate response. This project also streamlined the process of getting the SBML model from the BioModels repository and converting it into Antimony using tellurium. Overall, in this study, RAG was applied to biomodels to create a LLM assistant for researchers to help with modelling processes. 

The results of using the RAG-based approach allows users to understand the vast amount of information available in the BioModels repository in layman's terms to aid their modeling efforts. 

In order to test the fact that the RAG system shows reduced hallucination compared to the Llama3 output without RAG, fuzzy string matching using token set ratio was utilized to find the commonalities between the provided context and the final generated output. Finding the percentage of commonality between the context and final generated output can be used to find out whether the RAG system hallucinates - if there are significant semantic and word differences between the context and the final generated response, that could very likely mean that the model is hallucinating ideas and words that are not present in the context. 

To measure the hallucination and how off topic the non-RAG samples were compared to the RAG sample, cosine similarity was used to determine whether or not the non-RAG samples were off-topic and also were hallucinated answers \cite{cosine2}. Cosine similarity embeds the text and calculates the similarity between the two text vectors - the more similar they are, they less likely the non-RAG output is to be off topic, as the RAG output has been established as the non-hallucinated and on-topic baseline. 

Some example outputs are shown below. BIOMD000000054 and 
\newline 
BIOMD000000072 were analyzed together for the following queries.

\begin{tcolorbox}[colback=blue!5!white, colframe=blue!75!black, title=Original User Query: Example 1]
If I increase the concentration of ions in the reaction between ions and the energy where ions and the energy pool are reactants, will the output increase exponentially?
\end{tcolorbox}

\begin{tcolorbox}[colback=red!5!white, colframe=red!75!black, width=\linewidth, title=Llama3 Output Without RAG: Example 1, label=WITHOUTRAG1, breakable]
\section*{Assumptions and Clarifications}
I think there may be a few assumptions or misunderstandings in your question that need to be addressed. Firstly, the concept of ``energy pool'' is not a standard term in chemistry. It's possible that you're referring to the activation energy \(E_a\) required for a reaction to occur, but I'll assume it's something else entirely.

Secondly, increasing the concentration of ions is not necessarily related to the energy where ions are reactants. The two concepts are distinct: concentration refers to the number of particles per unit volume, while energy is a measure of the ability to do work.

\section*{Impact of Ion Concentration}
That being said, let's consider how increasing the concentration of ions might affect the reaction:

\begin{enumerate}
    \item \textbf{Concentration-dependent reactions}: If the reaction is concentration-dependent (i.e., the rate of reaction depends on the concentrations of reactants), then increasing the concentration of ions could lead to an increase in the reaction rate. This is because there are more particles available to collide and react.

    \item \textbf{Rate equations}: The reaction rate can be described by a rate equation, which takes into account factors like concentrations, activation energy, and other kinetic parameters. Increasing the concentration of ions might lead to a higher reaction rate, but this depends on the specific rate equation and the values of the kinetic parameters.
\end{enumerate}

However, exponential growth is unlikely in most chemical reactions. Chemical reactions typically follow power-law or logarithmic dependencies, rather than exponential ones. To illustrate this, consider a simple example:

\section*{The Arrhenius Equation}
The rate constant \(k\) for a reaction can be described by the Arrhenius equation:
\[
k = Ae^{-\frac{E_a}{RT}},
\]
where \(A\) is a frequency factor, \(E_a\) is the activation energy, \(R\) is the gas constant, and \(T\) is the temperature. Increasing the concentration of ions would not directly affect the exponential term in this equation.

\section*{Exponential Growth}
In general, exponential growth occurs when there are feedback loops or autocatalytic processes involved. Chemical reactions typically don't exhibit these types of feedback mechanisms, so exponential growth is less common in chemical reactions.

\section*{Conclusion}
To summarize: increasing the concentration of ions might lead to an increase in reaction rate, but this depends on the specific rate equation and kinetic parameters. Exponential growth is unlikely in most chemical reactions unless there are specific feedback loops or autocatalytic processes involved.

\end{tcolorbox}

\begin{tcolorbox}[colback=white, colframe=black, title=Llama3 Output With RAG: Example 1, label=WITHRAG1, breakable]
    \section*{Variables and Entities}
    \begin{itemize}
        \item $I$: Ions (millimole)
        \item $E$: Energy pool (millimole)
        \item $C$: Extracellular ion concentration (millimole)
        \item $p$: Membrane permeability (unknown unit)
        \item $W_2$: Ion pump activity (unknown unit)
        \item $T$: ATP (millimole)
        \item $M$: AMP (millimole)
        \item $A$: Adenylate pool (millimole)
        \item $U$: de novo AMP synthesis (unknown unit)
        \item $W$: AMP degradation (unknown unit)
        \item $n$: Dependence of AMP degradation on ATP (dimensionless)
        \item $m$: Dependence of AMP degradation on AMP (dimensionless)
    \end{itemize}

    \section*{Reactions}
    \begin{itemize}
        \item $U_1$: 
        \begin{itemize}
            \item $I$ is produced from an unknown input.
            \item Production rate is proportional to the product of $J$ and the cell's size ($P$).
        \end{itemize}
        
        \item $U_2$: 
        \begin{itemize}
            \item 3 units of $I$ react with $E$ to produce an unknown output.
            \item Production rate is proportional to:
            \begin{itemize}
                \item $I$: Concentration of ions
                \item $T$: Concentration of ATP
                \item $W_2$: Ion pump activity
            \end{itemize}
        \end{itemize}
    \end{itemize} 
    
    \section*{Important Note}
    The exact relationship between the production rate and the factors is unknown (could be linear, exponential, etc.).

    \section*{Conclusion}
    Based on the provided information, we \textbf{cannot definitively say} if increasing the concentration of ions (T) will lead to an exponential increase in the output of the reaction. We need more information on the specific form of proportionality between the production rate and the reaction.
\end{tcolorbox}

An interesting thing to note in \ref{WITHRAG1} is that the LLM with RAG likely did not hallucinate the answer. LLMs are notoriously prone to hallucination, however, with RAG, the LLM stated directly that it did not have enough information to answer the question \cite{juty2015}. In this study, we used token set ratio and fuzzy string matching to determine if the output included information not explicitly provided in the context, indicating that it would be hallucinating. However, there is still a level of expert knowledge in chemistry and chemical networks that is necessary for the user to have if they are to catch the hallucination in more advanced use-cases. 

However as one can see, the Llama3 output without RAG \ref{WITHOUTRAG1}, went off on a tangent and it did not include the context specific information (like reactions, and variable initializations that are essential to answer the questions). It also stated that energy pool is not a standard term in chemistry. While this may be true, it is a valid variable in the model and an entity that the user can ask questions about.

\begin{tcolorbox}[colback=blue!5!white, colframe=blue!75!black, title=Original User Query: Example 2, breakable]
Please give me the identity of the cell in a model involving ATP and Adenylate species: 
\end{tcolorbox}

\begin{tcolorbox}[colback=red!5!white, colframe=red!75!black, title=Llama3 Output Without RAG: Example 2, label=WITHOUTRAG2, breakable]
\section*{Overview of the Adenylate System}
The adenylate system is central to cellular metabolism, involving key molecules such as:

1. \textbf{ATP} (Adenosine Triphosphate): A high-energy molecule that stores energy in the form of phosphate bonds.
2. \textbf{ADP} (Adenosine Diphosphate): The product of ATP hydrolysis, releasing energy upon the removal of one phosphate group.
3. \textbf{AMP} (Adenosine Monophosphate): The product of ADP hydrolysis, resulting in a lower-energy molecule.

The primary reactions involved are:

\begin{itemize}
    \item \textbf{ATP synthesis}: \( \text{ATP} + \text{H}_2\text{O} \rightarrow \text{ADP} + \text{P}_i \) (phosphate)
    \item \textbf{ATP hydrolysis}: \( \text{ADP} + \text{P}_i \rightarrow \text{ATP} + \text{H}_2\text{O} \)
    \item \textbf{Phosphorylation}: \( \text{AMP} + \text{P}_i \rightarrow \text{ADP} \)
\end{itemize}

These reactions are crucial for energy storage, release, and regulation of cellular processes.

\section*{Energy Charge and Regulation}
To understand metabolic regulation, we can examine the \textbf{energy charge} of a cell, defined as:

\[
EC = \frac{[\text{ATP}] + 0.5[\text{ADP}]}{[\text{ATP}] + [\text{ADP}] + [\text{AMP}]}
\]

A high ATP concentration results in an energy charge close to 1, indicating sufficient energy, while a low energy charge prompts ATP production, activating pathways such as glycolysis and oxidative phosphorylation.

\section*{Adenylate Kinase Reaction}
The reaction:

\[
2 \text{ADP} \leftrightarrow \text{ATP} + \text{AMP}
\]

rapidly equilibrates the concentrations of ATP, ADP, and AMP, ensuring that energy demands are met promptly.

\section*{Conclusion}
In summary, the dynamics of ATP, ADP, and AMP concentrations are crucial for understanding how cells respond to changes in energy demands, highlighting the importance of the adenylate system in maintaining cellular homeostasis.
\end{tcolorbox}

\begin{tcolorbox}[colback=white, colframe=black, title=Llama3 Output With RAG: Example 2, label=WITHRAG2]
\section*{Cell Identity}
The identity of the cell being modeled is \textbf{"Erythrocyte"}. This classification is based on the following context:

\begin{itemize}
    \item \textbf{Cell Type}: This variable defines the type of cell being modeled, specifically, "Erythrocyte".
\end{itemize}

\end{tcolorbox}

In this example, the RAG output \ref{WITHRAG2} gave a clear, concise, and straight to the point answer, directly stating the identity of the cell that the user was looking for. By contrast, the Llama3 output without RAG \ref{WITHOUTRAG2} did not include any resemblance of a context specific answer in its response. It also started to include information about ADP and AMP, which were not asked for in the original user input.

\begin{tcolorbox}[colback=blue!5!white, colframe=blue!75!black, title=Original User Query: Example 3]
What is the assignment rule for ATP?
\end{tcolorbox}

\begin{tcolorbox}[colback=red!5!white, colframe=red!75!black, title=Llama3 Output Without RAG: Example 3, label=WITHOUTRAG3]
\section*{Assignment Rule for ATP (Adenosine Triphosphate)}
The assignment rule for ATP is a set of guidelines that determines how the phosphate groups are transferred from one molecule to another during cellular energy conversion.

\subsection*{Rules of the Assignment}
The assignment rule states that:

\begin{enumerate}
    \item The first phosphate group (\(\alpha\)-phosphate) is always transferred to ADP (Adenosine Diphosphate) to form ATP.
    \item The second phosphate group (\(\beta\)-phosphate) is always transferred to the next acceptor molecule, which is usually another enzyme or protein.
    \item The third phosphate group (\(\gamma\)-phosphate) is also transferred to the next acceptor molecule.
\end{enumerate}

In other words:

\begin{itemize}
    \item \(\alpha\)-phosphate \(\rightarrow\) ADP + \(\text{P}_i\)
    \item \(\beta\)-phosphate \(\rightarrow\) Next acceptor molecule
    \item \(\gamma\)-phosphate \(\rightarrow\) Next acceptor molecule
\end{itemize}

This assignment rule ensures that the energy stored in ATP is efficiently transferred to the cellular machinery, allowing cells to perform various functions such as muscle contraction, nerve impulses, and biosynthesis.

\end{tcolorbox}

\begin{tcolorbox}[colback=white, colframe=black, title=Llama3 Output With RAG: Example 3, label=WITHRAG3]
\section*{Assignment Rule for ATP}
The assignment rule for ATP, denoted as \( T \), is given by:

\[
T = \frac{A + 3E - \sqrt{6AE - 3E^2 + A^2}}{6}
\]

\subsection*{Derivation}
This formula is derived based on the following contexts:

\begin{enumerate}
    \item \textbf{Context 1}: This context lists \( T \) as the variable representing ATP.
    \item \textbf{Context 3}: This context provides the formula for calculating \( T \) using the variables \( A \) and \( E \) along with mathematical operations. Therefore, the provided formula represents the assignment rule for ATP (\( T \)).
\end{enumerate}
\end{tcolorbox}


For Example 1 \ref{WITHRAG1}, the token set ratio was 68, the second highest in the three examples provided in this paper. For Example 2 \ref{WITHRAG2}, the token set ratio is 69, the highest in the three examples. For Example 3 \ref{WITHRAG3}, the token set ratio is 61. In all three examples provided in this paper, the token set ratio was above 60, indicating it is likely that hallucination in these examples was reduced. While a perfect token set ratio would be 100, meaning that there is no difference between the given context and the final LLM output, that is unrealistic to achieve in an LLM-generated output. 

Along with the three examples explained here, the token set ratio was calculated for 6 other examples. The average of the token set ratio for these 9 examples was 52, which is lower than the samples included in the paper, but that could be due to the fact that there is additional information in the context due to the nature of segmenting (that may not be relevant to the user's question) compared to what is included in the final response.

In terms of cosine similarity, the average across all 9 samples was 0.335 when calculated between the RAG and non-RAG outputs. The established industry standard for cosine similarity is 0.5. The fact that the cosine similarity between the RAG and non-RAG outputs is less than 0.5, indicates that the non-RAG output is off topic compared to the RAG output. The RAG output has already been established as hallucination free so the cosine similarity calculations are used to determine whether the non-RAG output is on topic compared to the RAG output. The non-RAG output's poor similarity to the RAG output skews the overall cosine similarity calculation, highlighting that the non-RAG output is less effective in addressing the user's query.

\subsection{Implementation/Deployment}

In order to allow users who might not have an IDE set up to be able to use the RAG system, we also deployed the functionality on a website (\url{https://biomodelsrag.streamlit.app/}). The website provides the ability to choose specific biomodels from a drop down menu, visualize the model, and then enter a query. The previous conversation is also used as context.

The website was deployed as a Streamlit space \cite{streamlit}. The website also uses the same version of llama 3.1 (llama3.1-8b), but the instant version is used in the website, and the instruct version is used in the Python package. 

In the website, the \href{https://console.groq.com/docs/text-chat}{Groq Chat Completion API} was used, and in the Python package, a local deployment of ollama was used. 

\subsubsection{Selection of specific BioModels}

\begin{figure}[tbp]
    \centering
    \includegraphics[width=\textwidth]{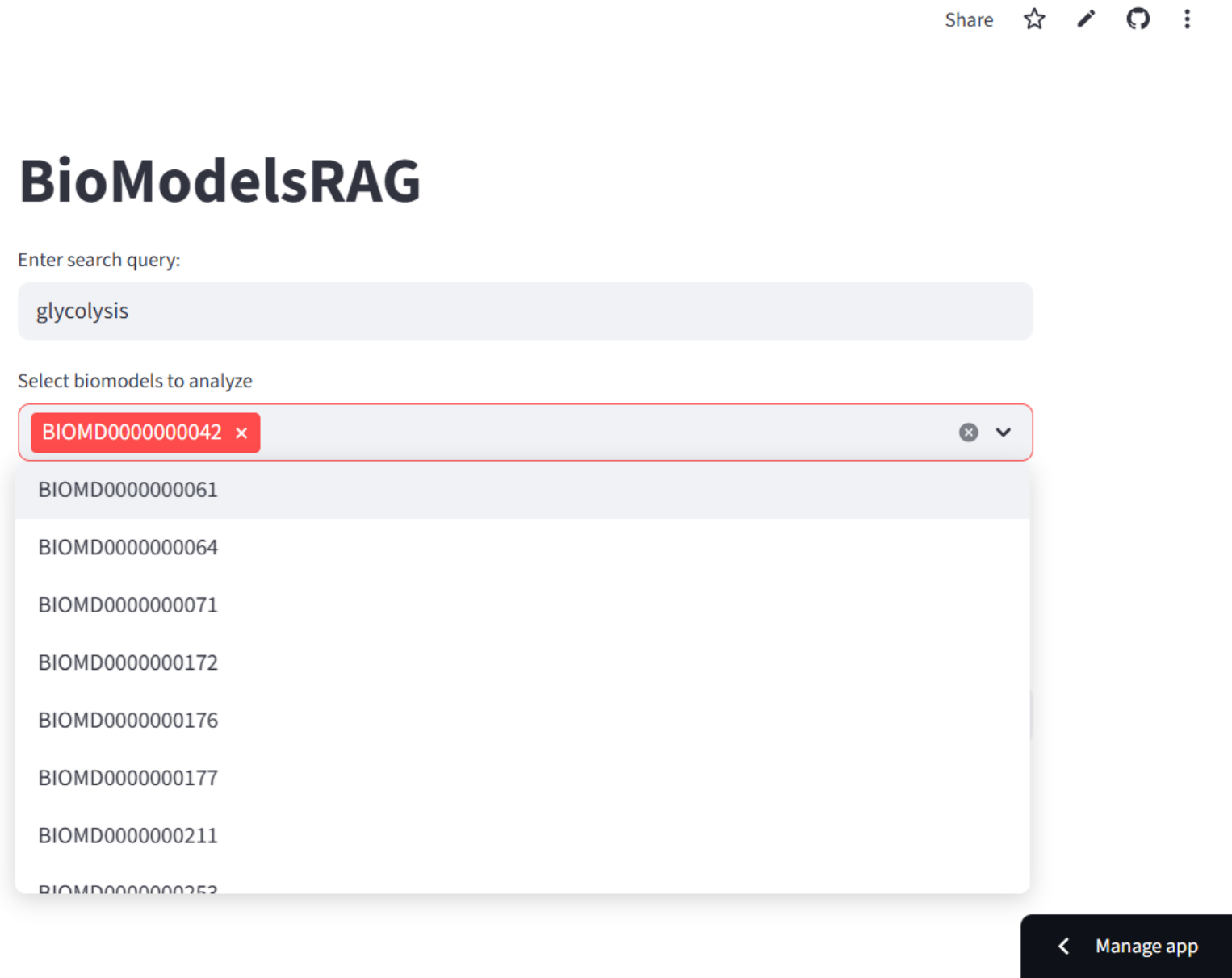}
    \caption{UI with dropdown for choosing specific biomodels to analyze. The user can enter a search query and then choosing multiple or single models from the dropdown. The dropdown will contain all of the models that matched with the query.}
    \label{fig:chosing}
\end{figure}

Both the website and the Python package implemented the ability to choose specific biomodels, as seen in Figure~\ref{fig:chosing}. The user is able to enter a search query, and choose a specific biomodel from the provided drop down. In the Python package, it is a command line interface where the user can enter yes or no to each model. 

\subsubsection{Conversation interface}

\begin{figure}[tbp]
    \centering
    \includegraphics[width=\textwidth]{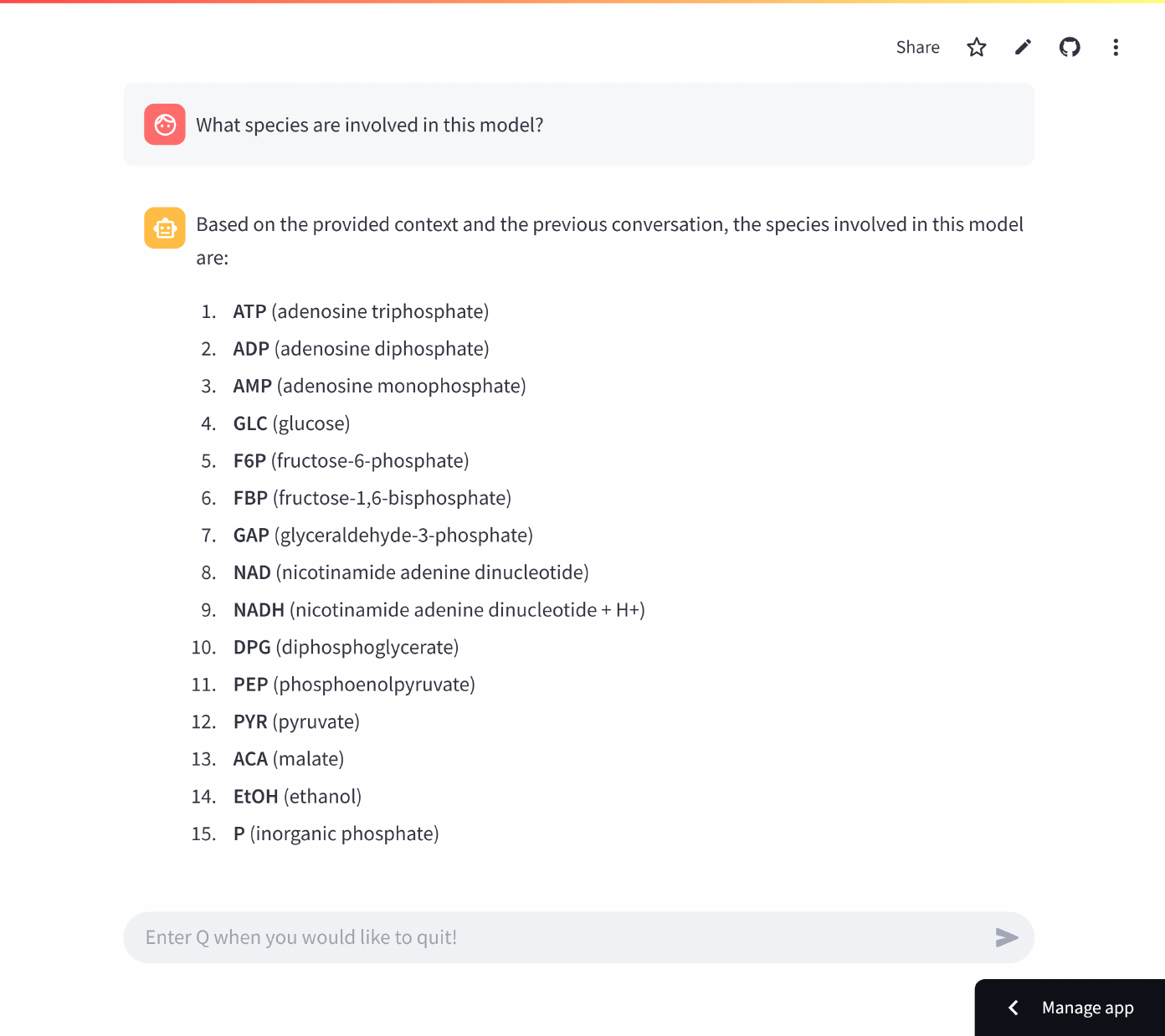}
    \caption{The UI for the chat portion is shown above. The user's query will be displayed with a grey background, and the LLM's output will be displayed right below it. A textbox will also be right below the LLM output so that the user can continue entering queries.}
    \label{fig:chat}
\end{figure}

Once the user has chosen which biomodels they would like to analyze, the user clicks `Analyze Biomodels'. Each biomodel is analyzed and added to the database, and a textbox appears where the user can enter their query. The query is be passed to the backend and the LLM provides the output. Once the frontend receives the output it displays it for the user to view. To quit, the user enters 'Q' to end the chat. A sample input and output is shown in Figure~\ref{fig:chat}.

\section{Conclusions and Implications}

The BioModelsRAG package and website presents an advancement in the text-based analysis and interaction with biological models. As stated by Clusmann et al., LLMs could have a significant impact on the field of medicine by aiding doctors in understanding and synthesizing information, and this can be extended to researchers as well in understanding the variety information available in BioModels \cite{clusmann}.

In conclusion, the pipeline described in this paper summarizes Antimony format models by creating a chatbot for users to better understand and analyze biomodels. To begin this workflow, a biomodels model is chosen by the user and is downloaded from a separate GitHub repository. The are converted from SBML into Antimony, segmented into specific sections and summarized using an LLM. The processed model is added to a vector database and is queried with the user's query to get the most accurate pieces of context for the LLM to use. This pipeline is a convenient way for users to investigate specific biomodels using a natural language interface. 

By leveraging Retrieval Augmented Generation (RAG) techniques, the package efficiently bridges the gap between complex biological models and user inquiries, serving as an effective assistant for researchers and educators \cite{louispatrick2024}. This integration enables researchers and practitioners to interact with biomodels in a dynamic, user-friendly manner, facilitating deeper insights and understanding of biological processes. In addition, BioModelsRAG has reduced hallucination as seen in the Query examples. Further avenues of research may include implementing GraphRAG to increase the relevance/accuracy of the output beyond baseline RAG. In all of the samples included in this paper, the token set ratio was above 60, indicating their reduced hallucination compared to the context they were based off of. However, for the non-RAG samples, their cosine similarity was 0.335, indicating their irrelevance and possible hallucination compared to the on-topic RAG output.

As the field of biological modeling continues to evolve, BioModelsRAG offers a robust, extensible framework for integrating advanced language processing capabilities, paving the way for future developments in AI-assisted biological analysis and interpretation.

\subsection{Documentation}
Documentation for the package is available at \url{https://thebobbob.github.io/BioModelsRAG/}. The website can be accessed at \url{https://biomodelsrag.streamlit.app/}.

\section*{Acknowledgments}
We thank Bartholomew E Jardine for assistance with the unit testing. HMS acknowledges that the research reported in this publication was supported by NIBIB of the National Institutes of Health under award number NIH grant number P41EB023912. BNK acknowledges receipt of a Mary Gates Award at UW, which helped support this work.

\section*{Author Contributions}
\textbf{Bhavyahshree Navaneetha Krishnan} - Formal analysis, Investigation, Methodology, Software, Visualization, Writing - original draft, Writing - review and editing, Funding acquisition, Validation
\newline 
\textbf{Adel Heydarabadipour} - Investigation, Methodology, Writing - review and editing, Software, Project administration, Supervision, Conceptualization 
\newline 
\textbf{Herbert Sauro} - Funding acquisition, Writing - review and editing, Conceptualization, Supervision
\bibliographystyle{elsarticle-num}
\bibliography{references}

\appendix

\section{Example 1}

\subsection{Prompt}
In the interaction between \( \text{IP3}_{\text{Cyt}} \) and \( \text{IP3}_{\text{PHGFP\_Cyt}} \), does the product increase monotonically?

\subsection{LLM Output}
A question about protein-ligand interactions!

IP3 (inositol 1,4,5-trisphosphate) is a ligand that binds to its receptor on the cytoplasmic side of the plasma membrane. The IP3-Cyt and IP3-PHGFP-Cyt proteins are both constructs of this receptor, with the difference being the presence of PHGFP (a photoactivatable GFP) in the latter.

To answer your question: yes, the product of the interaction between IP3 and either IP3-Cyt or IP3-PHGFP-Cyt increases monotonically. This is because the binding of IP3 to its receptor is a saturable process, meaning that at high concentrations of IP3, all available binding sites on the receptor are occupied.

In other words, as you increase the concentration of IP3, the amount of bound IP3 will initially increase rapidly, but then plateau or even level off as the receptor becomes fully saturated. This is often referred to as a "hyperbolic" relationship between ligand concentration and binding response.

So, to summarize: the product of the interaction between IP3 and either IP3-Cyt or IP3-PHGFP-Cyt increases monotonically with increasing concentrations of IP3, until the receptor becomes fully saturated.

\subsection{RAG Output}
\textbf{Identifying Relevant Information:}

\textbf{Context 3:} 
Defines variables related to \( \text{IP3}_{\text{Cyt}} \) and \( \text{IP3}_{\text{PHGFP\_Cyt}} \):
\begin{itemize}
    \item \( \text{kf\_IP3PH\_IP3\_PHGFP} \): Forward rate constant for the interaction
    \item \( \text{kr\_IP3PH\_IP3\_PHGFP} \): Reverse rate constant for the interaction
\end{itemize}

\textbf{Context 2:} 
Describes the reaction:
\begin{itemize}
    \item \( \text{IP3}_{\text{PHGFP}} \) (Display Name: IP3\_PHGFP) binds \( \text{IP3}_{\text{Cyt}} \) and \( \text{PH\_GFP\_Cyt} \) to form \( \text{IP3}_{\text{PHGFP\_Cyt}} \). 
    This reaction is influenced by the concentrations of \( \text{IP3}_{\text{Cyt}} \), \( \text{IP3}_{\text{PHGFP\_Cyt}} \), \( \text{KMOLE} \), \( \text{PH\_GFP\_Cyt} \), and the forward and reverse reaction constants, \( \text{kf\_IP3PH\_IP3\_PHGFP} \) and \( \text{kr\_IP3PH\_IP3\_PHGFP} \) respectively, multiplied by the concentration of the Cytosol compartment.
\end{itemize}

\textbf{Analyzing the Reaction:}
The reaction is a reversible binding process:
\[
\text{IP3}_{\text{Cyt}} + \text{PH\_GFP\_Cyt} \rightleftharpoons \text{IP3}_{\text{PHGFP\_Cyt}}
\]
The forward reaction is driven by \( \text{kf\_IP3PH\_IP3\_PHGFP} \), and the reverse reaction by \( \text{kr\_IP3PH\_IP3\_PHGFP} \).

\textbf{Determining Monotonic Increase:}
A monotonic increase means the product (\( \text{IP3}_{\text{PHGFP\_Cyt}} \)) continuously increases without any decrease.
However, since the reaction is reversible, the product formation will reach an equilibrium point where the rate of forward and reverse reactions becomes equal. The product \( \text{IP3}_{\text{PHGFP\_Cyt}} \) will not increase monotonically; it will reach a steady-state concentration at equilibrium.

\textbf{Answer:}

No, the product (\( \text{IP3}_{\text{PHGFP\_Cyt}} \)) does not increase monotonically. It reaches equilibrium where the forward and reverse reaction rates balance.

\subsection{Cosine Similarity}
0.34741276417104544

\section{Example 2}
\subsection{Prompt}
What species interacts with 16595 - \( \text{1D-myo-inositol 1,4,5-trisphosphate} \)?

\subsection{LLM Output}

\textbf{1D-myo-inositol 1,4,5-trisphosphate (IP3)} is a signaling molecule that plays a key role in various cellular processes. IP3 interacts with specific proteins to mediate its effects.

In particular, IP3 binds to and activates certain types of receptors known as IP3 receptors or type-2 inositol-receptor-like calcium-release channels. These receptors are found in the endoplasmic reticulum (ER) and other cellular compartments.

\textbf{IP3 receptor types include:}

\begin{itemize}
    \item \textbf{Inositol 1,4,5-trisphosphate receptor type 1 (IP3R1)}: Found in a variety of tissues, including brain, heart, and skeletal muscle.
    \item \textbf{Inositol 1,4,5-trisphosphate receptor type 2 (IP3R2)}: Expressed mainly in the brain and spinal cord.
    \item \textbf{Inositol 1,4,5-trisphosphate receptor type 3 (IP3R3)}: Found primarily in the testes.
\end{itemize}

When IP3 binds to its receptors, it triggers a cascade of events that ultimately leads to the release of calcium ions from the ER into the cytosol. This signaling pathway is important for various cellular functions, such as muscle contraction, cell proliferation, and neuronal function.

In summary, the species that interacts with 1D-myo-inositol 1,4,5-trisphosphate (IP3) are IP3 receptors (type-2 inositol-receptor-like calcium-release channels), specifically IP3R1, IP3R2, and IP3R3.

\subsection{RAG Output}
Based on the provided context, we can determine the following:
\begin{itemize}
    \item \textbf{IP3\_Cyt:} Represents inositol trisphosphate (IP3), which is a molecule that plays a crucial role in cellular signaling.
    \item \textbf{IP3\_PHGFP\_Cyt:} Represents a molecule that binds to a specific protein (PHGFP) and is attached to inositol trisphosphate (IP3).
\end{itemize}

Therefore, the species that interacts with 16595 - 1D-myo-inositol 1,4,5-trisphosphate (which is IP3) is \textbf{IP3\_PHGFP\_Cyt}. However, the context does not provide information on the specific interaction or the nature of the binding between IP3 and PHGFP.

\subsection{Cosine Similarity}
0.44768062992659813

\section{Example 3}

\subsection{Prompt}
What time unit does the mode involving adenylate take place under?

\subsection{LLM Output}
\textbf{A clever question!}

Glycolysis, the process of breaking down glucose into pyruvate, is a biochemical reaction that typically occurs in living organisms at the cellular level. The time scale for this process can vary depending on the specific context and the system being studied.

In general, glycolytic activity takes place over a relatively short period, often measured in minutes or seconds. Here are some examples:

\begin{itemize}
    \item In cells, glycolysis is an ongoing process that occurs continuously to provide energy for cellular activities. The time scale for this process can range from seconds to minutes.
    \item In muscle cells, glycolysis can occur rapidly in response to exercise or other energetic demands, taking place over a period of seconds to tens of seconds.
    \item In yeast and some bacteria, glycolytic activity is often studied in vitro (outside the living organism) and can take place over a timescale of minutes to hours.
\end{itemize}

So, to summarize, the time unit for a model involving glycolytic activity would typically be measured in seconds, minutes, or possibly hours, depending on the specific context and system being studied.

\subsection{RAG Output}
This model takes place under a time unit of \textbf{3600 seconds}. 

This information is found in the first piece of context under the \textbf{Unit definitions} section.

\subsection{Cosine Similarity}
0.3065796089259123

\section{Example 4}
\subsection{Prompt}
What is the variable initialization for glycolytic activity?

\subsection{LLM Output}
The glycolytic pathway is a series of enzyme-catalyzed reactions that convert glucose into pyruvate, producing energy in the form of ATP and NADH. The variables involved in glycolytic activity can be complex, but here are some key parameters that influence its regulation:

\begin{enumerate}
    \item \textbf{Glucose concentration:} The availability of glucose is a crucial factor in determining glycolytic activity. High glucose levels can stimulate glycolysis, while low glucose levels can inhibit it.
    \item \textbf{Insulin/insulin-like growth factor-1 (IGF-1):} Insulin and IGF-1 are key regulators of glycolytic activity. They activate the phosphatidylinositol 3-kinase (PI3K)/Akt signaling pathway, which phosphorylates and activates key enzymes in glycolysis.
    \item \textbf{AMP-activated protein kinase (AMPK):} AMPK is a master regulator of energy metabolism that can either stimulate or inhibit glycolytic activity, depending on the cellular context. It is activated by high levels of AMP (adenosine monophosphate), which indicates low energy levels.
    \item \textbf{Pyruvate kinase (PK) enzyme activity:} PK is the final enzyme in the glycolytic pathway, converting phosphoenolpyruvate to pyruvate. Its activity is often rate-limiting and regulated by various factors, including insulin/IGF-1 signaling, AMPK activation, and cellular energy status.
    \item \textbf{Pyruvate dehydrogenase kinase (PDK):} PDK phosphorylates and inactivates the pyruvate dehydrogenase complex (PDH), which is responsible for converting pyruvate to acetyl-CoA. PDK inhibition can stimulate glycolytic activity by allowing PDH to function.
    \item \textbf{Oxygen availability:} Glycolysis can be influenced by oxygen levels, as hypoxia can stimulate the production of glycolytic enzymes and facilitate glucose metabolism.
    \item \textbf{Cellular energy status:} The energy charge of the cell, reflected in ATP/ADP ratios, can also impact glycolytic activity. Cells with low energy reserves may prioritize glycolysis to generate more ATP.
    \item \textbf{Hormonal signals:} Hormones like epinephrine (adrenaline), norepinephrine (noradrenaline), and cortisol can regulate glycolytic activity by influencing the expression of key enzymes, such as hexokinase and phosphofructokinase 1.
\end{enumerate}

These variables interact with each other to determine the overall rate of glycolytic activity in a cell. Understanding their relationships is crucial for appreciating the complex regulation of glucose metabolism in different physiological contexts.

\subsection{RAG Output}
The variable initialization for glycolytic activity is \textbf{13.48}.

Here’s how we find that:
The second piece of context tells us that the variable \texttt{W3} corresponds to glycolytic activity.
The first piece of context lists the variable \texttt{W3} with an initialization of 13.48.

\subsection{Cosine Similarity}
0.3238045454480025

\section{Example 5}
\subsection{Prompt}
What is the variable initialization for KMOLE?

\subsection{LLM Output}
\textbf{KMOLE} stands for Kronecker-Mandelbrot, a model used in systems biology and computational biology. In the context of KMOLE, a variable initialization refers to the process of assigning initial values or conditions to variables that represent different biological components, such as genes, proteins, or metabolites.

In a KMOLE simulation, you typically need to specify the initial concentrations or activities of these variables to set up the system's dynamics. The specific initialization strategy depends on the problem you're trying to model and the available data.

Here are some common variable initialization strategies for KMOLE:

\begin{enumerate}
    \item \textbf{Experimental values:} Use measured values from experimental data, such as gene expression levels, protein concentrations, or metabolite abundances.
    \item \textbf{Literature-based values:} Utilize literature-derived values, like reported gene expression levels or protein abundance in similar systems.
    \item \textbf{Stoichiometric constraints:} Initialize variables based on stoichiometric relationships between components in the system, such as enzyme-substrate interactions.
    \item \textbf{Random initialization:} Assign random values to variables, often within a specific range or distribution (e.g., uniform or normal).
    \item \textbf{Parameter estimation:} Use optimization algorithms to estimate variable initializations by minimizing the difference between simulated and experimental data.
    \item \textbf{Model-based initialization:} Employ model-based approaches, such as using known regulatory relationships or gene regulatory networks (GRNs) to inform variable initializations.
\end{enumerate}

When initializing variables in KMOLE, it's essential to consider the following:

\begin{itemize}
    \item Ensure that the initialized values are physiologically plausible and within reasonable ranges.
    \item Verify that the chosen initialization strategy does not impact the overall behavior of the system or alter the model's predictive power.
    \item Consider incorporating uncertainty quantification techniques to account for inherent variability in biological systems.
\end{itemize}

Remember, the specific variable initialization approach depends on your research question, data availability, and modeling goals.

\subsection{RAG Output}
Based on the provided context, the variable initialization for KMOLE is \textbf{0.00166112956810631 (unit\_0)}.

\subsection{Cosine Similarity}
0.33455977313199053

\section{Example 6}

\subsection{Prompt}
What are the unit definitions for a model with \(\text{PIP2\_PHGFP\_PM}\)? Give me all of the unit definitions for this model.

\subsection{LLM Output}
\textbf{Unit Definitions for the \(\text{PIP2\_PHGFP\_PM}\) Model}

The \(\text{PIP2\_PHGFP\_PM}\) model is a membrane protein simulation model that simulates the behavior of phosphatidylinositol 4,5-bisphosphate (PIP2) and phagosomal GFP (PH-GFP) on a plasma membrane. Here are the unit definitions for this model:

\textbf{Length Units}
\begin{itemize}
    \item \textbf{nm}: nanometers (\(1 \, \text{nm} = 10^{-9} \, \text{meters}\))
\end{itemize}

\textbf{Energy Units}
\begin{itemize}
    \item \textbf{kJ/mol}: kilojoules per mole
    \item \textbf{kcal/mol}: kilocalories per mole
\end{itemize}

\textbf{Time Units}
\begin{itemize}
    \item \textbf{ms}: milliseconds (\(1 \, \text{ms} = 10^{-3} \, \text{seconds}\))
    \item \textbf{s}: seconds
\end{itemize}

\textbf{Concentration Units}
\begin{itemize}
    \item \textbf{mM}: millimolar (\(1 \, \text{mM} = 0.001 \, \text{mol/L}\))
\end{itemize}

\textbf{Other Units}
\begin{itemize}
    \item \textbf{pN}: piconewtons (\(1 \, \text{pN} = 10^{-12} \, \text{N}\), or newtons)
    \item \textbf{s\textsuperscript{-1}}: reciprocal seconds
    \item \textbf{mol/s}: moles per second
\end{itemize}

\subsection{RAG Output}
\textbf{Unit Definitions for the {\tt PIP2\_PHGFP\_PM} Model}

Based on the provided context, we can determine the unit definitions for the model with \(\text{PIP2\_PHGFP\_PM}\) as follows:

\textbf{Units}
\begin{itemize}
    \item \textbf{Length}: metre
    \item \textbf{Area}: \((1 \times 10^{-6} \, \text{metre})^2\)
    \item \textbf{Volume}: litre
    \item \textbf{Time Unit}: time\_unit (this appears to reference an existing unit, not define a new one)
\end{itemize}

\textbf{Substance}
\begin{itemize}
    \item \textbf{unit\_0}: \(1 \times 10^{-18}\) Avogadro
    \item \textbf{unit\_1}: \(\frac{\text{item}}{(1 \times 10^{-6} \, \text{metre})^2}\)
    \item \textbf{unit\_2}: \(\frac{1}{\text{second}}\)
    \item \textbf{unit\_3}: second
    \item \textbf{unit\_4}: \(1 / \text{dimensionless}^{-0}\)
    \item \textbf{unit\_5}: \(\frac{\text{metre}^3}{(1 \times 10^{-9} \, \text{dimensionless})^{-0} \times \text{mole second}}\)
    \item \textbf{unit\_6}: \(\frac{\text{item} \times \text{Avogadro}}{(1 \times 10^{-6} \, \text{dimensionless})^{-0} \times \text{litre}}\)
\end{itemize}

\textbf{Other Units}
\begin{itemize}
    \item \textbf{Extent}: substance
    \item \textbf{Substance per Volume}: \(\frac{\text{item}}{\text{litre}}\)
\end{itemize}

\subsection{Cosine Similarity}
0.15560592490167766

\end{document}